\documentclass[letterpaper]{article} 
\usepackage{aaai2026}  
\usepackage{times}  
\usepackage{helvet}  
\usepackage{courier}  
\usepackage[hyphens]{url}  
\usepackage{graphicx} 
\usepackage{natbib}  
\usepackage{caption} 
\usepackage{amssymb}  
\usepackage{lineno}
\usepackage{amsmath}

\usepackage{xcolor}

\title{Difficulty-Controlled Simplification of Piano Scores with Synthetic Data\\ for Inclusive Music Education}

\author{
Pedro Ramoneda\textsuperscript{\rm 1},
Emilia Parada-Cabaleiro\textsuperscript{\rm 2},
Dasaem Jeong\textsuperscript{\rm 3},
Xavier Serra\textsuperscript{\rm 1}
}

\affiliations{
\textsuperscript{\rm 1}Music Technology Group, Universitat Pompeu Fabra, Barcelona, Spain\\
\textsuperscript{\rm 2}Pedagogy Department, Nuremberg University of Music, Germany\\
\textsuperscript{\rm 3}Music \& Arts Learning Lab, Sogang University, Seoul, South Korea\\
\{pedro.ramoneda, xavier.serra\}@upf.edu,\ emiliaparada.cabaleiro@hfm-nuernberg.de,\ dasaem.jeong@sogang.ac.kr
\vspace{-0.5cm}
}

\definecolor{darkgreen}{RGB}{0, 128, 0}
\definecolor{darkred}{RGB}{139, 0, 0}

\usepackage{booktabs, makecell}

\usepackage{enumitem}

\newcommand{\eg}{e.\,g., }
\newcommand{\ie}{i.\,e., }

\usepackage{tikz}
\usetikzlibrary{shapes.geometric, arrows}

\tikzstyle{startstop} = [rectangle, rounded corners, minimum width=3cm, minimum height=1cm,text centered, draw=black, fill=red!30]
\tikzstyle{process} = [rectangle, minimum width=3cm, minimum height=1cm, text centered, draw=black, fill=blue!30]
\tikzstyle{arrow} = [thick,->,>=stealth]

\usepackage{bibentry}

\begin{document}

\maketitle

\begin{abstract}
Despite its potential, AI advances in music education are hindered by proprietary systems that limit the democratization of technology in this domain.  
In particular, AI-driven music difficulty adjustment is especially promising, as simplifying complex pieces can make music education more inclusive and accessible to learners of all ages and contexts. Nevertheless, recent efforts have relied on proprietary datasets, which prevents the research community from reproducing, comparing, or extending the current state of the art. In addition, while these generative methods offer great potential, most of them use the MIDI format, which, unlike others, such as MusicXML, lacks readability and layout information, thereby limiting their practical use for human performers. This work introduces a transformer-based method for adjusting the difficulty of MusicXML piano scores. Unlike previous methods, which rely on annotated datasets, we propose a synthetic dataset composed of pairs of piano scores ordered by estimated difficulty, with each pair comprising a more challenging and easier arrangement of the same piece. We generate these pairs by creating variations conditioned on the same melody and harmony and leverage pretrained models to assess difficulty and style, ensuring appropriate pairing. The experimental results illustrate the validity of the proposed approach, showing accurate control of playability and target difficulty, as highlighted through qualitative and quantitative evaluations. In contrast to previous work, we openly release all resources (code, dataset, and models), ensuring reproducibility while fostering open-source innovation to help bridge the digital divide.
\end{abstract}


\section{Introduction}

\begin{figure}[ht]
    \centering
\includegraphics[width=1\linewidth]{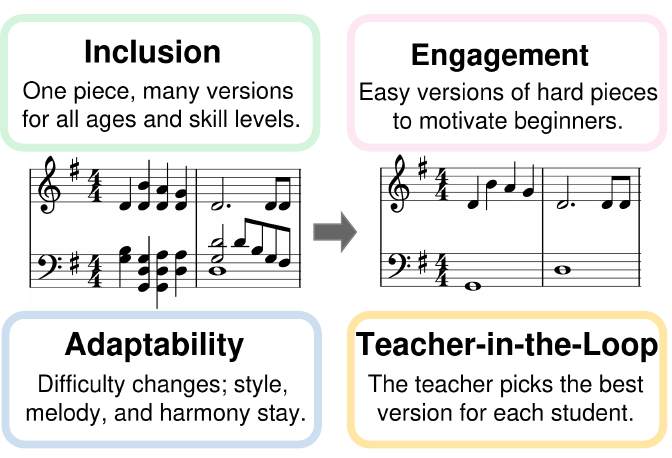}
    \vspace{-0.6cm}
    \caption{Accessible Music through Difficulty-Aware Generation. One musical idea, adapted to each player. AI difficulty control enables everyone to participate and learn.}
    \label{fig:teaser}
    \vspace{-0.3cm}
\end{figure}

In general, education systems have traditionally marginalized certain subjects, such as music. This is particularly clear within STEAM\footnote{Science, Technology, Engineering, Arts, and Mathematics} approaches, where music is often instrumentalized rather than playing a central role \cite{liao2016interdisciplinary}. However, technological advances such as AI also had the potential to empower disadvantaged subjects, \eg by fostering more inclusive learning opportunities and enabling teachers to engage in posthuman pedagogies \cite{cook2016posthuman}.  
Yet, current AI applications in music education often rely on closed proprietary systems, which hinders reproducibility \cite{haim2023open} while undermining  progress and democratization of technology in the education sector; thus, widening the digital divide \cite{miao_guidance_2023}.

\begin{figure*}[ht]
    \centering
    \includegraphics[width=1\linewidth]{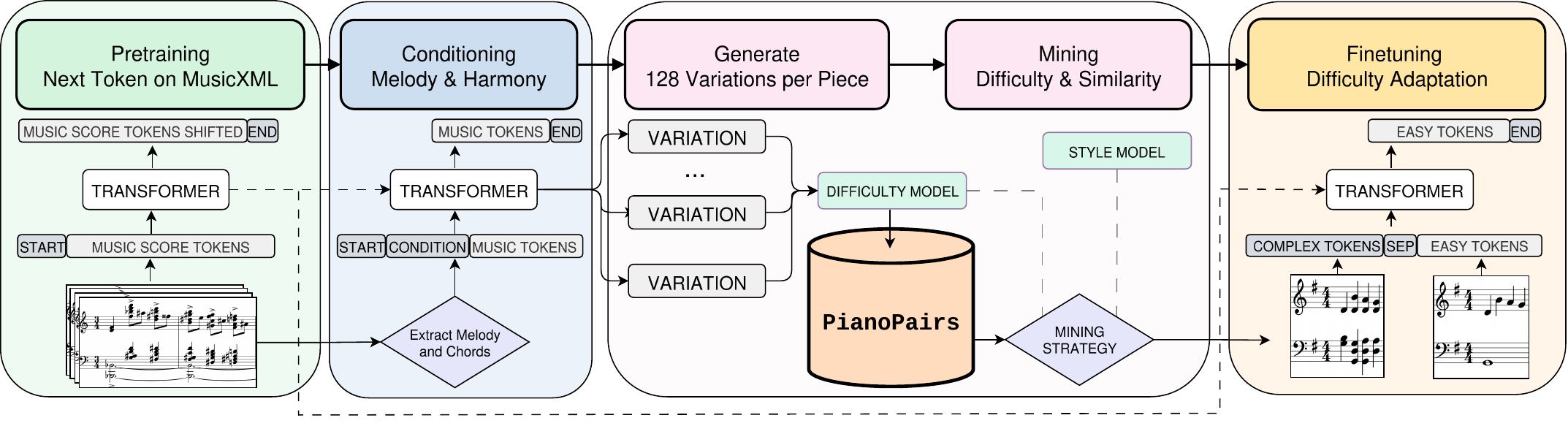}
    \vspace{-0.6cm}
    \caption{Proposed Training approach for Adaptative Score Generation.}
    \label{fig:main-architecture}
    \vspace{-0.3cm}
\end{figure*}







One promising technological aid that can meaningfully support educational approaches to music learning is controlling the difficulty of generated musical scores. In particular for piano, one of the most typically learned musical instruments, this approach introduces transformative possibilities (see Figure~\ref{fig:teaser}),
as it can make compositions more accessible and adaptable to the needs of individual learners \citep{corozine2015arranging,cacavas1975music,sanganeria2024tuning}. Beginners could play simplified versions of pieces that would usually take years of practice to master, letting them enjoy famous or difficult works. This fosters motivation as 
it allows playing iconic pieces early in the learning journey, \ie the idea behind collections of simplified music that are occasionally used in education \citep{heumann2017eine,heumann2017best,marks2009easy,catania2023sixteen}. Similarly, even if not so common for pedagogical purposes, the creation of intricate arrangements tailored to high-profile performances can facilitate adapting repertoires for special events; thus, expanding the creative potential of generative systems~\citep{zhu2020pop,wang2020pop909}.

Difficulty control offers the possibility of enhancing traditional pedagogies by allowing teachers to progressively increase the complexity of the pieces as the student's proficiency improves, providing a more engaging and personalized learning experience. 
It is important to mention, that such an adaptive framework is understood in this work as part of a Teacher-in-the-Loop approach aimed to support the teaching-learning process without impairing the social dimension of learning~\cite{wieczorek2025ai}, something often ignored by commercial AI-based music education systems, which typically pursue autodidactic learning.

Although promising, this line of research has been limited by the absence of public datasets containing multiple difficulty levels for the same piece, whose production
is extremely costly
due to the need of expert annotation and careful alignment.
Indeed, although major actors
such as Yamaha~\citep{Suzuki2023}, \ie a global leader in music education, and Simply Piano~\citep{gover2022music}, \ie the only unicorn startup in the field, have explored difficulty-aware generation, 
their works remain closed source, making reproducibility and further research nearly impracticable. 
In addition, even though large language models (LLMs) have proven to be useful in many creative tasks, they still struggle with symbolic music due to the lack of large-scale public datasets and the unique structural properties of music~\citep{yuan2024chatmusician,zeng2021musicbert}. 
This justifies the development of smaller, specialized models that better capture the hierarchical and multi-voice nature of symbolic music~\citep{huang2018music,le2025survey}. 
Therefore, we aim to advance open-source music education technologies as a way to increase AI democratization and reduce the digital divide by: 
\begin{enumerate}[label=(\roman*)]
\item Modeling the task of difficulty-aware score simplification as a conditional generation problem, where the goal is to “translate” existing piano scores from one difficulty level to another while preserving their musical structure \ie melody, harmony, and style. This formulation balances simplifying the musical discourse while maintaining fidelity, something we achieved through  
a training methodology for a transformer model that does not rely on expert-annotated data. Figure 2 sketches our full methodology pipeline.

\item Generating 128 difficulty-controlled variations for more than 3\,000 pieces conditioned on fixed melody and harmony.  The resulting synthetic dataset, \texttt{PianoPairs}, is released in MusicXML format, which not only preserves layout and readability, but also guarantees playability, something essential for real educational uses.

\item Introducing a mining strategy that identifies musically similar score pairs with different difficulty levels by combining pretrained models for style similarity and difficulty estimation~\citep{ramoneda2024explainable,wu2025clamp3}. 

\item Publicly releasing the dataset, models, and code,\footnote{All data and code will be released upon acceptance.} together with extensive evaluations and demo\footnote{Demo available at: \url{https://pramoneda.github.io/diff2diff/}} that demonstrate their value in real music education and arrangement scenarios.
\end{enumerate}

\subsection{Related Work}

Over the past decade, deep learning and transformer-based architectures have driven progress in automatic music generation~\citep{vaswani2017attention}. To simplify data processing, MIDI representations have become the standard due to their compactness and simplicity~\citep{huang2018music}. However, using MIDI presents several limitations, particularly in terms of readability, which makes this type of data unsuitable in some practical applications, such as music education~\citep{Ramoneda2025SightReading} or classical composition~\citep{Ramoneda2024RefinPaint}. 
Indeed, unlike formats such as MusicXML, which provides detailed symbolic representations necessary for human readability and interpretation, MIDI often lacks essential nuances for performance, such as neglecting the voicing of music~\citep{foscarin2024cluster}. This is especially problematic for difficulty-aware score generation, whose value lies in producing not only playable but also readable output for musicians. 
Inspired by work on optical music recognition -- OMR~\citep{Zaragoza2020}, we adopted their encoding scheme for music notation~\citep{olimpic2024} in order to tackle these challenges.


Although music simplification is inspired by other simplification tasks in NLP~\citep{siddharthan2014survey}, symbolic music presents a grammar whose complexity exceeds by far the one of natural language, besides the fact that large datasets are not available; thus, limiting the direct transfer of such methods.
Concerning the task at hand, \citet{Suzuki2023} presents a highly relevant approach for the presented work. Nevertheless, while also using a pair dataset for training when exploring automatic difficulty modification of piano scores, the underlying data are not shared, making reproducibility unfeasible. 
A similar limitation affects the dataset of \citet{gover2022music}, collected from the Simply Piano app, which is proprietary and does not contemplate any type of release, making it impossible to verify independently.  
Furthermore, creating a similar dataset would raise legal concerns, as it would involve generating arrangements of varying difficulty, often stemming from the commercial scores composed in the 20th century; thus, still protected under copyright laws.
In contrast, other approaches  \cite{Terao2023,Nakamura2018} 
focus on score reduction, \ie transforming multi-instrument works into solo piano arrangements. These methods describe difficulty using correlated descriptors, such as note density or hand span requirements, rather than directly addressing difficulty as a conceptual entity. Although 
offering insight into the arrangement of playable scores, they lack the granularity needed for precise difficulty control.




Despite the fact that  organizing educational material in steps of increasing difficulty (where mastering each step unlocks the next) is a prerequisite to learn effectively, objective grading of music in the field of instrumental learning is largely unexplored compared to other fields, such as learning reading \cite{wareham1967development}. 
In addition, learning to play an instrument and performing
is more than just reading what is written on the score; it also involves body movement, sound, and principles related to specific  interpretation schools, among other aspects~\citep{cook1999analysing}.
Thus, performance difficulty could be defined as the effort required by a pianist to satisfy the expressive, technical, and notational demands of a passage~\citep{ramoneda2024}.  

Due to a lack of objective criteria, the difficulty is open to a degree of subjectivity, which leads to disagreements between individual educators and classification systems, and makes formalization still an important task~\cite{wareham1967development}.
Currently, the rankings of music boards, conservatories, composers, and publishers, despite the variability that accompany them, generally reveal global trends.
This study adopts the expert‑annotated difficulty labels from the dataset by~\citet{ramoneda2024}, 
which follow the nine‑level grading scale by Henle Verlag and have supported recent models for piano difficulty estimation and playability~\citep{zhang2023symbolic,ramoneda2024explainable,ramoneda2023sheetdifficulty}. 

\section{Methodology}

Figure~\ref{fig:main-architecture} shows our framework for generating piano scores with controlled difficulty. We start by pretraining a model on a large corpus of MusicXML files using next-token prediction to learn the symbolic music structure. Using MusicXML, we preserve the layout and readability of the scores, making the generated output suitable for real-world educational settings. After pretraining, the model is finetuned with melodic and harmonic conditioning to ensure that generated variations retain the original musical essence.

In contrast to melody and harmony, we do not condition directly on difficulty using a prepended token. There are two main reasons: (i) We follow previous approaches for the task at hand, which rely on paired score datasets~\cite{Suzuki2023,gover2022music}, inspired by the literature on neural machine translation---as these data sets are private, we design our own synthetic dataset; 
(ii) Since simplifying a score is not just about reducing difficulty---among other aspect,  it must remain stylistically consistent---we control how stylistically similar the pair elements are, instead of conditioning directly on difficulty using a prepended token. 

Finally, we generate 128 variations for each of more than 3\,000 original pieces, all sharing melody and harmony. Using pretrained models for difficulty~\citep{ramoneda2024} and style similarity~\citep{wu2025clamp3}, we mine variation pairs that are musically consistent but vary in difficulty. These pairs form the basis for training a difficulty-aware model to generate musically coherent scores at a target difficulty.

\subsection{Architecture and Pretraining}

Our architecture is a transformer-based, decoder-only model designed to generate piano scores while preserving musical structure. We were inspired by the findings of~\citet{liu24ce} 
for designing efficient sub-billion-parameter models. 
The training process begins with causal language modeling, where the objective is to predict the next token in a sequence based on the previous ones. Similar autoregressive transformers, inspired by text transformers~\cite{vaswani2017attention}, have already shown to work well for symbolic music~\cite{huang2018music,huang2020pop,thickstun2024anticipatory}.  \vspace{0.2cm}

\noindent
Formally, given a sequence of tokens,
\[
X = \{x_1, x_2, \ldots, x_T\},
\]
the model is trained to maximize the conditional probability;
\[
P(x_t \mid x_{<t}),
\]
by minimizing the cross-entropy loss:
\[
\mathcal{L} = -\frac{1}{T} \sum_{t=1}^{T} \log P(x_t \mid x_{<t}).
\]
This enables the model to learn coherent musical structures in an autoregressive manner.

\begin{figure}[t]
    \centering
    \includegraphics[width=0.99\linewidth]{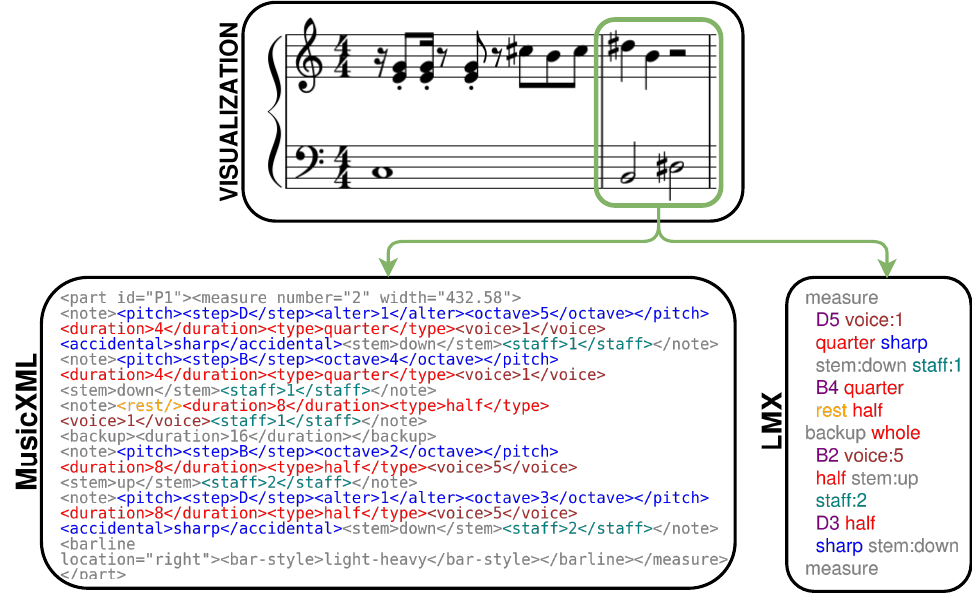}
    \vspace{-0.1cm}
    \caption{Comparison of input representations. Above: two measures  of a piano score. 
    Below: 
    MusicXML  (left);  Linearized MusicXML -- LMX (right) token sequence. 
    }
    \label{fig:lmx}
    \vspace{-0.3cm}
\end{figure}

\subsubsection{Input Representation}
\label{sec:input-repr}

We adopt \textit{Linearized MusicXML} (LMX), introduced by \cite{olimpic2024}, as input representation. LMX converts the hierarchical partwise structure of MusicXML into a single token sequence obtained through a depth-first traversal of the XML tree. Musical elements, \eg pitch, duration, dynamics, etc., become discrete tokens, while notes from different parts (right and left hands in piano scores) appear in chronological order. This design keeps the rich descriptive power of MusicXML, but removes excessive verbosity, making the data directly compatible with transformer architectures and enabling efficient training.
Figure~\ref{fig:lmx} presents an example measure: the left column shows the original MusicXML (1711 characters), and the right column shows the corresponding LMX encoding (173 characters or 23 tokens). This illustrates how LMX preserves detail while providing a concise and sequence-friendly format.

LMX offers clear advantages over other symbolic representations such as Humdrum’s kern** format, which although well structured, has fewer available scores and cannot represent some piano voice interactions. LilyPond instead focuses on typing rather than describing notation, and its scripting syntax complicates machine learning workflows. MIDI lacks visual layout, which is critical for educational applications. In contrast, MusicXML is the most common interchange format, enjoys robust editor support, and is aligned with standard OMR pipelines; LMX leverages these benefits while eliminating XML’s syntactic rigidity.

In summary, LMX provides a practical compromise: it retains the widespread adoption and semantic richness of MusicXML, but presents the data as a flat sequence suited for generation tasks such as score simplification.

\subsection{Conditioning on Melody and Harmony}

To guide generation according to musical context, we prepend a conditioning prefix \( C \) to the input sequence \( X \). This prefix can contain either symbolic tokens (e.g., melodic skyline) or continuous embeddings (e.g., pitch profiles). The resulting sequence is:
\[
X' = \{C, x_1, x_2, \ldots, x_T\},
\]
which is used as causal input by the transformer.

During training, we prevent the model from directly copying the conditioning tokens into its predictions by applying a masking scheme. Let the mask be defined as
\[
M = \{m_1, m_2, \ldots, m_{|C|+T}\}, \quad m_t = 
\begin{cases}
1 & \text{if } t \leq |C| \\
0 & \text{otherwise}
\end{cases}
\]
so that the loss is only computed over the unmasked tokens:
\[
\mathcal{L} = -\frac{1}{T} \sum_{t=1}^{T} (1 - m_t) \log P(x_t \mid x_{<t}).
\]

This allows the model to benefit from contextual cues in \( C \) without learning to replicate them. In what follows, we describe the two conditioning signals used in our system.

\subsubsection{Melody Skyline Conditioning}

We extract the melodic contour using the \textit{melody skyline} algorithm~\cite{uitdenbogerd1999melodic}—a simple but effective technique selecting the highest pitch at each time step in a polyphonic score.

Let \( A_t \) be the set of active pitches in time step \( t \). The skyline pitch is then defined as
\[
s_t = \max_{p \in A_t} p.
\]
This gives a symbolic sequence \( C = \{s_1, s_2, \ldots, s_T\} \), which sketches the melodic outline of the piece. We tokenize this sequence and prepend it to the original input. The melody skyline remains a competitive baseline for symbolic melody extraction~\citep{simionetta2019convolutional}, and its efficiency makes it ideal for multiple experiments~\citep{hu2024musically}.

\subsubsection{Harmony Conditioning}

To encode harmonic context, we compute a pitch class profile---a normalized histogram over the 12 pitch classes:
\[
\mathbf{p} = (p_1, \ldots, p_{12}) \in \mathbb{R}^{12}, \quad \sum_i p_i = 1, \quad p_i \geq 0.
\]
This vector is projected in the model embedding space and positioned on the input as part of \( C \).  Previous work shows that harmonic profiles improve global coherence in symbolic generation~\citep{genchel2019explicitly,morpheus}. 

To encourage diversity during generation, we perturb \( \mathbf{p} \) with controlled uniform noise:
\[
\tilde{p}_i = \mathrm{clip}(p_i + \epsilon_i, 0, 1), \quad \epsilon_i \sim \mathcal{U}(-0.2 p_i, 0.2 p_i),
\]
followed by normalization:
\[
\tilde{\mathbf{p}} \leftarrow \frac{\tilde{\mathbf{p}}}{\sum_i \tilde{p}_i}.
\]
This results in harmonic conditioning that is expressive, yet not overly deterministic, inspired by previous works in text~\cite{lachaux2020target}.

\begin{figure}[t!]
    \centering
    \includegraphics[width=0.99\linewidth]{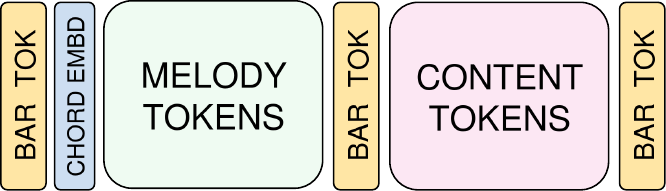}
     \vspace{-0.1cm}
    \caption{Input representation for training the model conditioned on melody and harmony. The input sequence consists of conditioning tokens (melody and harmony) followed by the score tokens.}
    \label{fig:input}
    \vspace{-0.6cm}
\end{figure}

\subsection{Generation of the \texttt{PianoPairs} Synthetic Dataset}

To enable difficulty conditioning during training, we construct a synthetic dataset, \texttt{PianoPairs}, containing multiple versions of each piano piece, all sharing the same melodic and harmonic structure but differing in difficulty.

For each original piece, we sample \( N = 128 \) score variations using our autoregressive transformer model, conditioned on both the melody skyline and the pitch profile of the original score (see Figure~\ref{fig:input}). This results in stylistically coherent sequences anchored in the same musical context.

To estimate the difficulty of each generated score, we adapt the approach proposed in previous work~\citep{ramoneda2024explainable}, which introduced a small neural network model (often referred to as RubricNet) trained on 12 handcrafted features extracted from symbolic scores. Instead of using the original architecture, we employ a calibrated Gaussian Naive Bayes classifier~\citep{john1995estimating}.

The classifier outputs a probability distribution over difficulty levels \( \mathcal{Y} \). The predicted difficulty \( \hat{y} \) is obtained via:
\[
\hat{y} = \arg\max_{y \in \mathcal{Y}} \; P(y \mid \mathbf{f}).
\]
We also retain the maximum probability \( \max_y P(y \mid \mathbf{f}) \) as a confidence estimate and apply a threshold to discard samples with uncertain predictions, removing the least confident 25\% of the variations. This results in a clean and structured pool of variations labeled by difficulty.

To ensure that these difficulty-controlled variations remain stylistically faithful to their original piece, we evaluate their similarity using a pretrained CLaMP-3 model~\citep{wu2025clamp3}. Trained through contrastive learning across multiple modalities, CLaMP-3 produces embeddings that capture high-level stylistic information.

Let \( \mathbf{e}_{\text{v}_1} \) and \( \mathbf{e}_{\text{v}_2} \) be the CLaMP-3 embeddings of two generated variations of the same original piece, where \( \text{v}_2 \) has a lower difficulty than \( \text{v}_1 \). The stylistic similarity between them is computed using cosine similarity:
\[
\text{sim}(\mathbf{e}_{\text{v}_1}, \mathbf{e}_{\text{v}_2}) = \frac{\mathbf{e}_{\text{v}_1} \cdot \mathbf{e}_{\text{v}_2}}{\|\mathbf{e}_{\text{v}_1}\| \, \|\mathbf{e}_{\text{v}_2}\|}.
\]
For each original piece and each difficulty level, we consider all possible pairs of generated variations that differ in difficulty. Among these, we retain only the most stylistically similar 50\% ones, according to the cosine similarity. This filtering step ensures that the final dataset consists of difficulty-controlled variation pairs that preserve the musical character of the underlying piece.

\subsection{Difficulty Adaptation}

To enable control over the difficulty of generated scores, we use the difficulty-labeled pairs previously introduced as \texttt{PianoPairs}. Each pair consists of two musically similar versions of the same piece with different difficulty levels. These are used to train the model to transform a score into an easier version while preserving its musical identity.

\subsubsection{Supervised Conditioning}

We frame difficulty adaptation as a sequence-to-sequence task using supervised learning. Given a pair of scores \( (X_{\text{hard}}, X_{\text{easy}}) \), where the second is a simplified version of the first, we concatenate them with a difficulty control token in between. Specifically, the input to the model is:
\[
X' = \{x_1, \ldots, x_T, \texttt{[SEP]}, y_1, \ldots, y_S\},
\]
where \( \{x_t\}_{t=1}^{T} = X_{\text{hard}} \), \( \{y_s\}_{s=1}^{S} = X_{\text{easy}} \), and \texttt{[SEP]} is a special token that explicitly indicates the intended reduction in difficulty.

We treat the first part of the sequence as a prompt and mask its contribution in the loss calculation. Formally, the loss is defined over the second segment:
\[
\mathcal{L} = -\frac{1}{S} \sum_{s=1}^{S} \log P(y_s \mid x_{1:T}, \texttt{[SEP]}, y_{<s}).
\]
This encourages the model to generate simplified versions of the input conditioned on the original score and the difficulty control token.






\section{Experimental Setup}
\subsubsection{Initial (real-world) corpus}

We pretrain on piano scores using the \textit{XMander dataset} derived from the popular website Musescore~\cite{xmader_2022_musescore_dataset}, which we downloaded for this work in 2023. The corpus is divided into 80\% for training and 20\% for validation. After selecting only piano scores with  
two staves and considering copyright restrictions,\footnote{
In addition, our research is covered by Directive (EU) 2019/790, which allows text and data mining on content protected by copyright when the work is done for scientific research.} we end up with 136k scores for training, \ie 450M of tokens in total. This corpus is used to pretrain the language model that captures symbolic structure and layout and serves as a foundation to subsequent
conditioning on melody and harmony, and later in the difficulty pairs.

\subsubsection{\texttt{PianoPairs} (synthetic) corpus}

We selected 3\,500 pieces from the validation split of the
real-world corpus to serve as bases, extracting their melody and harmony. For each piece, we generated 128 variations to subsequently estimate their difficulty and construct difficulty-aware pairwise comparisons. This results in 5.3M pairs with a difficulty gap of 1 level or more and 2.3M with a gap of 2 or more. 
We refer to pairs retrieved without any data mining strategy as \textbf{\textit{random}}. As an intelligent mining strategy of training with \texttt{PianoPairs} dataset, we discarded 25\% pieces with low-confidence difficulty labels using a Gaussian Naive Bayes classifier and filtered out 50\% of the least similar pairs per piece, as detailed in the methodology. After filtering, we obtained 1.1M training pairs and 231k validation pairs for a gap of 1, and 395k training pairs and 72k validation pairs for a gap of 2. We used these training pairs to finetune the final conditioning model to reduce the difficulty, referring to the mining strategy behind them as \textbf{\textit{filtered}}. Finally, by computing the mean cosine distance between CLaMP embeddings for each pair, we observe that the filtered mining strategy yields lower average distances than the random one. For a gap of 1, the mean distance decreases from 0.296 (random) to 0.237 (filtered), and for a gap of 2, from 0.296 to 0.246. This suggests that filtering selects semantically closer pairs, especially for smaller difficulty gaps.

\subsubsection{Training setup}
We train three transformer-based models (30 layers, hidden size 2048, embedding size 512, vocabulary size 512, 8 attention heads, and dropout 0.1; 95M total parameters). All models use Rotary Positional Embeddings -- RoPE~\cite{su2021roformer}, FlashAttention~\cite{dao2023flashattention2}, and a KV cache for inference. The base model is pretrained unconditionally with a causal language modeling objective on LMX sequences. We then finetuned two additional variants with LoRA adapters~\cite{hu2022lora} considering rank 64 and alpha 128: the second model is conditioned on melody and harmony, and the third model conditions the input difficulty followed by a \texttt{<sep>} token and a lower target difficulty, using sequences of up to 8\,000 tokens (instead of a 4\,000-token context in prior models). Due to budget constraints, all models are trained with an effective batch size of 4 (by accumulation of gradients when needed), a learning rate of $6\times10^{-4}$, and the \texttt{AdamW} optimizer on RTX~2080~Ti GPUs. All experiments were carried out with a fixed random seed (42) to ensure reproducibility. We acknowledge that a broader search for hyperparameters could further improve performance.


\begin{table}[t!]
\centering
\small
\renewcommand{\arraystretch}{1.2}
\begin{tabular}{cc|cccc}
\toprule
\textbf{M. Strat. } & \textbf{Gap} & \textcolor{darkgreen}{$\downarrow$} & $\sim$ & \textcolor{darkred}{$\uparrow$} & \textbf{Dis.} \\
\midrule
random &  1 & 68.8\% & 24.4\% & 6.8\%  & .309 \\
filtered &  1 & 71.8\% & 22.4\% & \textbf{5.8\%}  & \textbf{.283} \\
random  &  2 & 66.6\% & 25.0\% & 8.5\%  & .305 \\
filtered &  2 & \textbf{74.4\%} & \textbf{17.1\%} & 8.4\% & .299 \\
\bottomrule
\end{tabular}
\vspace{-0.2cm}
\caption{Original benchmark results on different mining strategies (M. Strat.).  Arrows show if the variation is easier \textcolor{darkgreen}{$\downarrow$}, harder \textcolor{darkred}{$\uparrow$}, or similar ($\sim$) to the original. Dis is the CLaMP cosine distance.}
\label{tab:global-outcome}
\vspace{-0.3cm}
\end{table}

\subsubsection{Evaluation framework}

We manually selected 52 copyrighted pieces composed after 2023 for avoiding leakage with \texttt{PianoPairs}: 10 pop, 10 rock, 10 k-pop, 10 latin, 10 film arrangements, and 2 classical (newly discovered works by Mozart and Chopin). For each piece, we generated 20 variations to evaluate whether difficulty can be reduced while melody and harmony are preserved. This set will be referred to as \textbf{\textit{original benchmark}}.  
Since most pieces are less than level~2, which limits difficulty diversity, 
we create the \textbf{\textit{extended benchmark}} using synthetic pieces generated by conditioning on melody and harmony, as done in the synthetic dataset. The final benchmark comprises 96 pieces on four difficulty levels: level~1~(35), level~2~(29), level~3~(28), and level~4~(4); while spaning genres: k-pop~(21), rock~(19), latin~(18), pop~(17), film~(16), and classical~(5). 
Finally, we generate 10 variations for each song to evaluate our system.

\section{Objective experiments}

Tables~\ref{tab:global-outcome} and~\ref{tab:global-outcome-ext} summarize the performance of our system on the \textit{original benchmark} and the \textit{extended benchmark}, respectively. Each table compares two data mining strategies: \textit{random}, which uses all available pairs with different difficulty levels without filtering, and \textit{filtered}, which selects pairs based on confidence and stylistic similarity, 
in two difficulty gaps, \ie gap 1, gap of one level or more than 1, and gap 2, gap of two levels or more. For each setting, we report the proportion of generated variations that are easier ($\downarrow$), similar ($\sim$) or harder ($\uparrow$) than the original, as well as
the mean cosine distance between CLaMP embeddings \textit{(Dis.)} for each pair, the original piece, and the generated one. 



\begin{table}[t!]
\centering
\small
\renewcommand{\arraystretch}{1.2}
\begin{tabular}{cc|cccc}
\toprule
\textbf{M. Strat} & \textbf{Gap} & \textcolor{darkgreen}{$\downarrow$} & $\sim$ & \textcolor{darkred}{$\uparrow$} & \textbf{Dis.} \\
\midrule
random   &  1 &  76.8\% & 19.4\% & 3.9\%  & .269\\
filtered  &  1 & 76.3\% & 19.3\% & 4.6\% & \textbf{.252} \\
random   &  2 &  72.1\%  & 23.2\%  & 4.7\%  & .315 \\
filtered  &  2 & \textbf{78.4\%} & \textbf{18.6\%} & \textbf{3.0\% } & .317 \\
\bottomrule
\end{tabular}
\vspace{-0.2cm}
\caption{Original extended results on different mining strategies (M. Strat.). Arrows show if the variation is easier \textcolor{darkgreen}{$\downarrow$}, harder \textcolor{darkred}{$\uparrow$}, or similar ($\sim$) to the original. Dis is the CLaMP cosine distance.}
\label{tab:global-outcome-ext}
\vspace{-0.6cm}
\end{table}

\begin{figure}
    \centering
\includegraphics[width=0.7\linewidth]{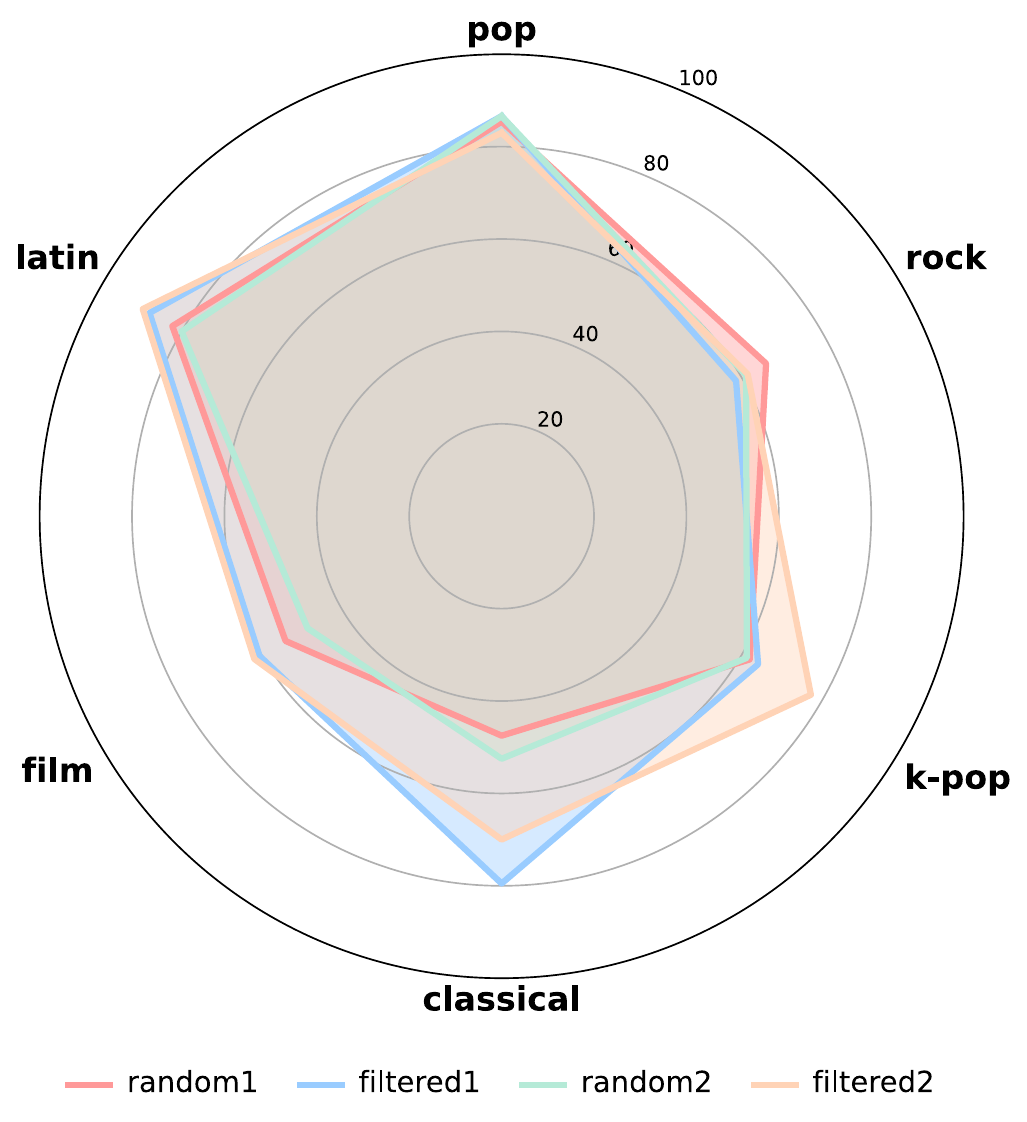}
    \vspace{-0.2cm}
    \caption{Percentage of \(\downarrow\) (easier) adaptations per genre for each experiment on the original benchmark.  Axis represent  genres; values indicate the percentage of correct adaptations.}
    \label{fig:spider-original}
    \vspace{-0.3cm}
\end{figure}

In the original benchmark (cf.\ Table~\ref{tab:global-outcome}), all mining strategies show a strong simplification trend, with more than 65\% of the variations being easier than the original. The \textit{filtered} strategy improves this trend further, reaching 74.4\% for a difficulty gap of 2, and also reduces the percentage of harder variants to 5.8\% (gap 1) and 8.4\% (gap 2). Furthermore, the mean cosine distance between the CLaMP embeddings decreases in all \textit{filtered} settings, suggesting that the generated variations stay closer in style and content to the original. In contrast, increasing the difficulty gap tends to increase the stylistic distance between pieces. 

In the extended benchmark (cf.\ Table~\ref{tab:global-outcome-ext}), simplification rates are even higher, reaching 78.4\% for the \textit{filtered} strategy with gap 2. The percentage of harder outputs also drops to just 3.0\% in this case. Although the cosine distance remains similar or slightly higher for large gaps, the overall trend confirms that filtering based on semantic similarity improves control over difficulty adaptation while maintaining musical coherence. In particular, the \textit{ filtered} strategy consistently outperforms the \textit{random} baseline across all configurations in terms of both simplification accuracy and stylistic proximity.

Figure~\ref{fig:spider-original} shows the percentage of successful downward simplifications per genre in the original benchmark, which is genre-balanced. The results are generally consistent across genres, except in classical music (where the \textit{filtered} strategy
dominates for gap 1) and k-pop (where it dominates for gap 2). This suggests that the \textit{filtered} strategy 
better captures genre-specific characteristics, leading to more effective simplifications in certain styles.

\begin{figure}
    \centering
\includegraphics[width=0.83\linewidth]{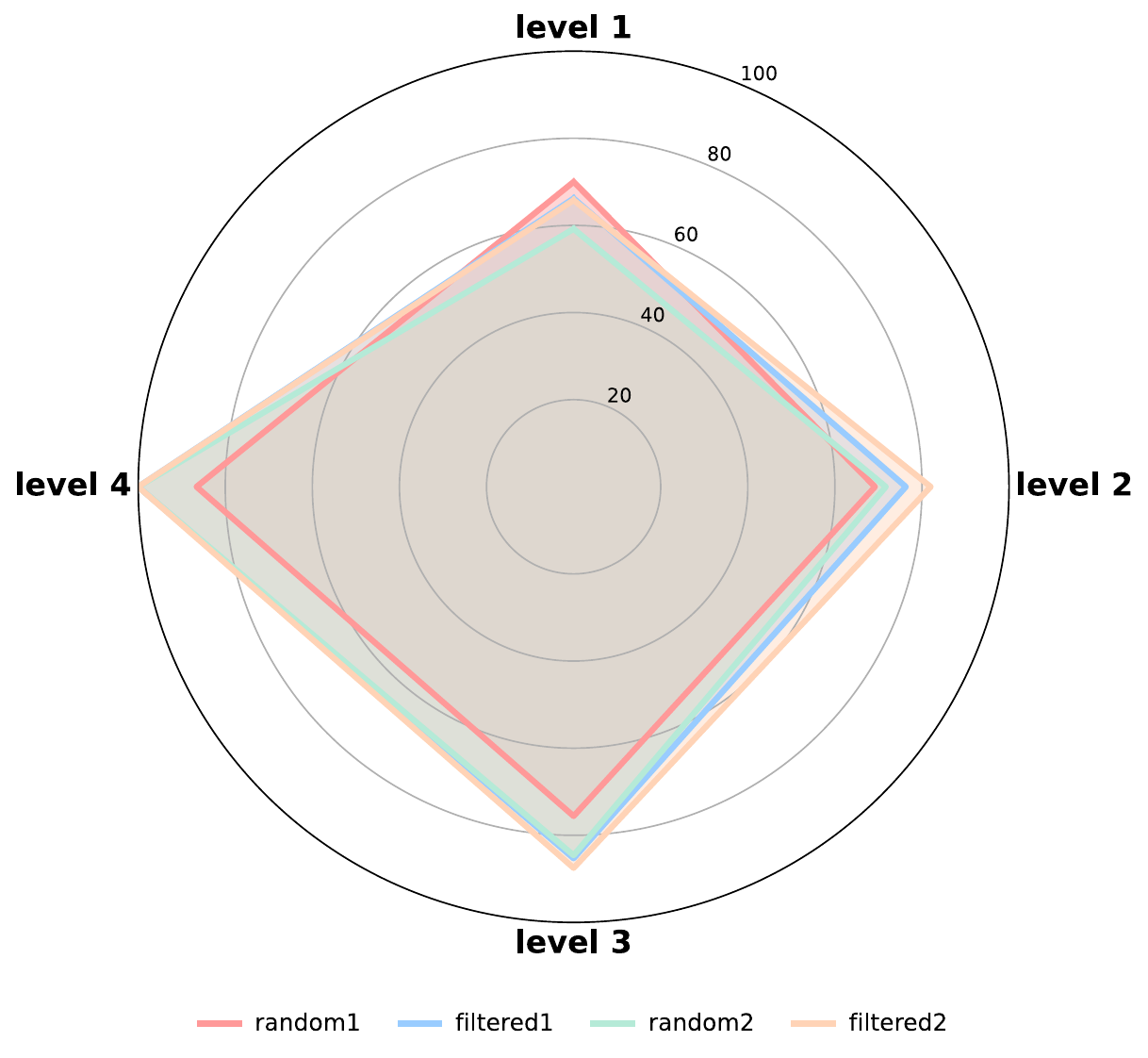}
    \vspace{-0.2cm}
    \caption{Percentage of \(\downarrow\) (easier) adaptations per difficulty level (from 1 to 4) for each experiment on the extended benchmark. 
    }
    \vspace{-0.3cm}
    \label{fig:spider-levels-extended}
\end{figure}

Finally, Figure~\ref{fig:spider-levels-extended} analyzes the extended benchmark by difficulty level, illustrating the performance in a wider range of skills. The \textit{filtered} data mining strategy performs best in almost all levels. In addition, a clear trend emerges: It is generally easier to reduce the difficulty of harder pieces than of simple ones. This aligns with the intuition that more complex scores offer more opportunities for simplification.

\section{Subjective experiments}


In order to assess the usability and expected impact of the proposed approach in real-world scenarios,  we conducted a perceptual test with 10 expert Spanish pianists, all of whom have at least a \textit{Bachelor of Music in Piano Performance} and are active piano teachers  (age: $\mu = 40.1$ years; playing piano experience: $\mu = 30$ years),
to evaluate the quality of our generation of difficulty-controlled scores.  
All responses were anonymously collected, the procedure conducted adhered to the ethical guidelines of the Declaration of Helsinki, and consent to evaluate the collected responses for research purposes was obtained from all involved participants. 


For each of the six selected and original musical excerpts, participants reviewed two simplified versions based on models trained on filtered and random mining strategies. These were presented to the participants in random order to avoid bias. 
Participants are invited to select which of the two proposed simplified versions relates the best to the following  aspects:
\textit{easiness}, \textit{coherence/original style preservation}, \textit{readability}, \textit{playability}, and \textit{overall preference}. 
The subjective test also included three questions to be rated on a Likert scale from 1 to 5 aimed at measuring: (i) instructional control (``Would this system help you better control the teaching process?'' ); (ii) pedagogical alignment (``Could the generated versions realistically fit established pedagogical criteria?''); (iii) personalized instruction  (``Would this system support personalization and student motivation?''). 

\begin{figure}[t!]
  \centering
  \includegraphics[
    width=\columnwidth
  ]{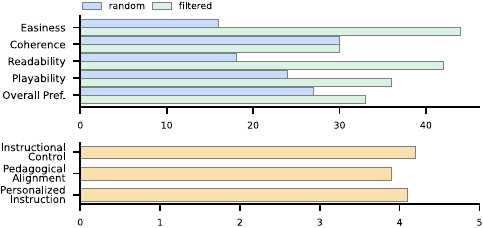}
  \vspace{-0.5cm}
      \caption{Subjective evaluation: votes per criterion (top, random vs.\ filtered) and average pedagogical ratings (bottom).}
  \label{fig:votes-scores-filtered-random}
  \vspace{-0.5cm}
\end{figure}

The results in Figure~\ref{fig:votes-scores-filtered-random} show a strong overall preference for the filtered model in all criteria except coherence, where both versions were rated equally. We computed the \(p\)-values with an exact two-sided binomial test (\(H_0: p=0.5\)). In particular, filtered output was significantly favored for easiness (44 vs.\ 16, (\textit{p-val} $=0.0004$)) and readability (42 vs.\ 18, \textit{p-val} $=0.0027$), while the difference in playability (36 vs.\ 24, \textit{p-val} $=0.1550$) and global preference (33 vs.\ 27, \textit{p-val} $=0.5190$) was numerically better but not statistically significant. This suggests that the proposed model successfully simplifies scores while maintaining readability and playability. Ratings on the Likert scale questions were also high, with average scores of 4.20 (std \(=\) 1.25) for instructional control, 3.90 (std \(=\) 1.55) for pedagogical alignment, and 4.10 (std \(=\) 1.22) for personalized instruction. This indicates an overall perceived pedagogical usefulness, despite individual differences.

\section{Concluding Remarks}


Our approach investigates the effectiveness of
mining strategies and conditioning mechanisms to simplify piano scores while preserving their style. 
This work should be contextualized within a
Teacher-in-the-Loop framework, enabling music educators to adapt the repertoire to student needs in order to promote more inclusive and motivating learning experiences. 
Although automation in education may raise concerns, we see this system as a possible pedagogical aid.

Our results confirm that the proposed mining strategy improves control and coherence, outperforming random pairing in all settings. 
Despite promising results, it is worth mentioning that the method focuses on arrangements characterized by a clear melody–accompaniment structure, which is common in many musical genres and forms, but could limit its generalization capabilities, as it might perform worse
on highly polyphonic or contrapuntal textures.
Another limitation is that simplification primarily affects accompaniment, while melodic lines are rarely simplified, partly due to the absence of explicit melodic complexity control. 
In addition, even though generation of upward difficulty
is relevant for advanced learners, this was not explored in our study. Finally, the evaluation relies on synthetic pairs due to the lack of public-aligned datasets, which limits direct benchmarking against human arrangements. Still, the model effectively reduces difficulty and provides a solid base for further research on a challenging task, though additional work remains necessary to refine how musical character and nuance are preserved.
All in all, our system enables generating piano score simplifications aligned with real educational goals by going beyond the autodidactic paradigms that currently dominate educational technologies in the music domain. 
By releasing our models, data, and evaluation framework, we lay the basis for reproducible research in difficulty-aware music generation; thus,
opening new paths towards democratization of cutting-edge technologies within music education.





\appendix

\bibliography{aaai2026}

\end{document}